\documentclass[onecolumn,useAMS,usenatbib]{mn2e}

\usepackage{color}
\usepackage{amsmath}
\usepackage{graphicx}
\usepackage{amssymb}
\usepackage{psfrag}

\newcommand{\Bav}{\bar{B}}
\newcommand{\eps}{\epsilon}
\newcommand{\Aph}{A_\phi}
\newcommand{\beq}{\begin{equation}}
\newcommand{\eeq}{\end{equation}}
\newcommand{\bB}{{\bf B}}
\newcommand{\bj}{{\bf j}}
\newcommand{\be}{{\bf e}}
\newcommand{\br}{{\bf r}}
\newcommand{\pom}{\varpi}
\newcommand{\beqy}{\begin{eqnarray}}
\newcommand{\eeqy}{\end{eqnarray}}
\newcommand{\td}[2]{\frac{d{#1}}{d{#2}}}
\newcommand{\pd}[2]{\frac{\partial{#1}}{\partial{#2}}}
\newcommand{\brac}[1]{\left({#1}\right)}
\newcommand{\emag}{\mathcal{E}_{mag}}
\newcommand{\epol}{\mathcal{E}_{pol}}
\newcommand{\etor}{\mathcal{E}_{tor}}
\newcommand{\nn}{\nonumber}
\newcommand{\curl}{\nabla\times}
\renewcommand{\div}{\nabla\cdot}
\newcommand{\GSu}{\Delta_*\! u}
\renewcommand{\lor}{\boldsymbol{\mathcal{L}}}

\title[Magnetic fields in axisymmetric NSs]{Magnetic fields in axisymmetric neutron stars}
\author[S. K. Lander and D. I. Jones]
       {S. K. Lander$^{1}$\thanks{skl@soton.ac.uk} and
        D. I. Jones$^{1}$\thanks{d.i.jones@soton.ac.uk}\\
$^{1}$University of Southampton, Southampton, U. K.}
\begin{document}


\pagerange{\pageref{firstpage}--\pageref{lastpage}} \pubyear{0000}

\maketitle

\label{firstpage}

\begin{abstract}
We derive general equations for axisymmetric Newtonian MHD and use these as the
basis of a code for calculating equilibrium configurations of rotating
magnetised neutron stars in a stationary state. We investigate the
field configurations that result from our formalism, which include
purely poloidal, purely toroidal and mixed fields. For the mixed-field
formalism the toroidal component appears to be bounded at less than
7\%. We calculate distortions induced both by magnetic fields and by
rotation. From our non-linear work we are able to look at the realm of
validity of perturbative work: we find for our results that
perturbative-regime formulae for magnetic distortions agree to within
10\% of the nonlinear results if the ellipticity is less than $0.15$ or
the average field strength is less than $10^{17}$ G. We also
consider how magnetised equilibrium structures vary for different
polytropic indices.
\end{abstract}


\section{Introduction}

The physics of neutron stars classes them among the most extreme objects in
the known Universe: their densities, rotation rates and magnetic
fields are all among the highest known for any astrophysical
object. Typical neutron star magnetic fields are up to $\sim
10^{12-13}$ G, whilst for magnetars this figure is $\sim 10^{15}$
G. Since magnetic fields induce a distortion in a star, a rotating
magnetised neutron star could be a significant source of gravitational
radiation. With the advent of second-generation gravitational wave
detectors like Advanced LIGO, we may soon be in a position to
observe neutron stars through their gravitational radiation signals
--- and hence have a new probe of the physics of these stars.

Understanding magnetic distortions requires an understanding of the
neutron star's interior field; NSs with relatively weak exterior
fields could still have significant ellipticities if
they have a much stronger field in their bulk. Here we model a NS as an
infinitely conducting polytropic fluid and examine various kinds
of magnetic field: purely poloidal, purely toroidal and mixed-field
configurations. The numerical scheme we use is able to deal with
extremely strong fields and fast rotation, so we are able to study the
theoretical properties of very highly magnetised stars, as well as
examining how well perturbative results hold away from the weak-field
regime.

It has long been predicted that magnetic fields will distort a fluid
star \citep{chandfermi}. It was found that this
distortion only becomes appreciable if the magnetic energy $\emag$ of
the star is comparable with its gravitational energy $W$; since
neutron stars have tremendous self-gravity it follows that one would
only expect very strong magnetic fields to generate any significant
distortion. This suggests that one would expect magnetars to be the
most distorted NSs and hence of most interest to gravitational wave
astronomy (with the caveat that this early work is for an
incompressible fluid and so is of limited relevance to NSs). For the
results presented in this paper we will quantify this statement by
scaling our code-generated results to real neutron star values.

A number of studies of magnetically deformed stars exist. These have
included work focussed on poloidal, toroidal or mixed fields, and
boundary conditions where the fields either vanish on the surface of
the star or decay at infinity. Changing any of these can lead to very
different results, so the uncertainty we have about the geometry of NS
magnetic fields translates into an uncertainty about how distorted
neutron stars are.

Analytic approaches have been restricted to weak fields and small
deformations, as the nonlinear nature of stronger magnetic fields
rapidly makes the problem intractable. Early work treated deformations
of incompressible fluids (see, e.g.,
\citet{ferraro,roberts,ostrgunn}), a
simplifying assumption but not terribly physical for real stars. The
first studies of compressible stars assumed very simplistic density
distributions and magnetic fields confined within the
star \citep{woltjer,wentzel}; later \citet{goossens} treated the
problem of a poloidal field matched to an external dipole, extending
the work of \citet{ferraro}. More recently, work has focussed on
the problem of magnetic deformations related specifically to neutron
stars \citep{haskell}, including a mixed-field case with vanishing
exterior field.

In addition to analytic work, a number of studies have used numerical
methods to calculate magnetic distortions. \citet{monaghan} and
\citet{roxburgh} calculated field geometries and surface
distortions for various polytropes, allowing for an exterior magnetic
field. Their work was perturbative and so restricted to weak
fields. More recently a second-order perturbation technique has been
applied for the strong fields found in magnetars \citep{ioka}. Other
studies of highly magnetised stars have solved the fully non-linear
problem, to allow for more highly deformed configurations than could
be accurately determined using a perturbative approach. This was
originally done for strong magnetic fields confined within the
star \citep{ostrhart}, by extending an earlier self-consistent field
method for rapidly-rotating stars \citep{ostrmark}. For purely poloidal
fields an improved numerical method was devised which enabled the
calculation of highly distorted equilibrium
configurations \citep{miketinac}; it was found that for very strong
fields the maximum density of the star could move away from the centre
to make the geometry of the density distribution toroidal. Solutions
have also been found using a mixed-field
formalism \citep{tomieri}. Finally, relativistic effects have been
considered: fully relativistic solutions for purely poloidal
fields \citep{bocquet} and purely toroidal fields \citep{KKSY} and
partially-relativistic solutions in the mixed-field case
\citep{KKKK,colaiuda}. In the Discussion we shall return to the role
of boundary conditions in the mixed-field case.

This paper is a study of the various stationary, axisymmetric
equilibrium solutions for Newtonian fluid stars in perfect MHD. We
show that the full equations of MHD reduce under these limits to two
general cases: a mixed-field case (which includes purely poloidal
fields as a special case) and purely toroidal fields. The mixed-field
formalism dates back to \citet{gradrubin} and was recently used by
\citet{tomieri} to study mixed-field stars. We are not aware of any
previous work using the other, purely toroidal, case for Newtonian
MHD. In the mixed-field case the toroidal fields vanish outside the
star, but the poloidal fields only decay at infinity; we consider this
boundary condition more realistic than the condition of zero exterior
fields used by much of the previous work discussed above. With our
formalism, we investigate the resulting field configurations,
including the relative strengths that toroidal and poloidal fields can
have, and the maximum theoretical field strength a fluid star can have
whilst remaining in an axisymmetric stationary equilibrium state. We
also look at distortions induced by magnetic fields, including the
effect of changing the polytropic index. We examine the validity of
perturbative results for magnetic distortions in the strong-field
regime. Finally, we rescale all our code results to canonical neutron
star values and ensure we are always comparing magnetic and rotational
effects in the same physical model star.

\section{Axisymmetric formalism}

\subsection{Governing equations}

We model a rotating magnetic neutron star by assuming that it is in a
stationary state, axisymmetric with both the magnetic dipole axis and
the spin axis aligned, and comprised of infinitely conducting material
(the perfect MHD approximation). We work in electromagnetic units. The
equations that describe this system are the Euler equation
\beq \label{newton2}
-\frac{1}{\rho}\nabla P-\nabla\Phi_g+\nabla\Phi_r
+\frac{\lor}{\rho}=0,
\eeq
together with Poisson's equation
\beq \label{diffpoisson}
\Delta\Phi_g = 4\pi G\rho,
\eeq
Amp\`{e}re's law
\beq
\nabla\times\bB=4\pi\bj
\eeq
and the solenoidal constraint
\beq
\nabla\cdot\bB=0.
\eeq
We close the system of equations by assuming a barotropic equation of
state:
\beq
P=P(\rho).
\eeq

In the above equations $P,\rho,\Phi_g,\Phi_r,{\bf{j}},{\bf{B}}, G$ and
$\lor$ are the pressure, density, gravitational potential, centrifugal
potential, current density, magnetic field, gravitational constant and
Lorentz force ($\lor={\bf j\times B}$), respectively.

Although the formalism allows for different choices of the centrifugal
potential $\Phi_r$ and equation of state $P=P(\rho)$, we will work
with a rigidly rotating star:
\beq
\Phi_r = \frac{\Omega^2_0\pom^2}{2},
\eeq
where the angular velocity $\Omega_0$ is a constant and $\pom$ the
cylindrical polar radius; and a polytropic
equation of state:
\beq \label{polyEOS}
P=k\rho^{1+1/N},
\eeq
where $k$ is some constant and $N$ the polytropic index.

The assumption of axisymmetry simplifies the equations
considerably. Taking the curl of equation \eqref{newton2} leaves us
with the requirement that
\beq \label{zerocurl}
\curl\brac{\frac{\lor}{\rho}} = 0.
\eeq
Additionally, the solenoidal nature of ${\bf B}$ allows us to write it
in terms of some streamfunction $u$, defined through the relations
\beq \label{streamfn}
B_\pom = -\frac{1}{\pom}\pd{u}{z}\ ,\ B_z = \frac{1}{\pom}\pd{u}{\pom}.
\eeq
One may also define a solenoidal ${\bf B}$-field by using the vector
potential ${\bf A}$, where ${\bf B = \curl A}$; we will use the
$\phi$-component $\Aph$ later. These two definitions are related by
$u=\pom\Aph$.
We also define a differential operator $\Delta_*$ by
\beq
\Delta_*\equiv \pd{^2}{\pom^2}-\frac{1}{\pom}\pd{ }{\pom}+\pd{^2}{z^2}.
\eeq

Using the two conditions \eqref{zerocurl} and \eqref{streamfn}, one
can show that Amp\`{e}re's law in axisymmetry may be rewritten as
\beq \label{jandBgen}
4\pi\bj = \frac{1}{\pom}\nabla(\pom B_\phi)\times\be_\phi
          -\frac{1}{\pom}\GSu\ \be_\phi
\eeq
 --- see section \ref{genmagderivation} for a
full derivation.

\subsection{Mixed-field formalism}

In axisymmetric perfect MHD with mixed poloidal and toroidal fields,
the magnetic field and current are related through the Grad-Shafranov
equation (see, e.g., \citet{gradrubin} or section \ref{mix_formal}):
\beq
4\pi\rho\td{M}{u} = -\frac{1}{\pom^2} \brac{\GSu 
                                       +f(u)\td{f}{u}},
\eeq
where $f(u)\equiv \pom B_\phi$ and $M(u)$ is defined through
$\nabla M(u)\equiv\lor/\rho$. Combining the Grad-Shafranov equation with
\eqref{jandBgen} yields
\beq
\bj = \frac{1}{4\pi}\td{f}{u}\bB + \rho\pom\td{M}{u}\be_\phi.
\eeq
Finally we use the notation of \citet{tomieri} and
\citet{chandprend}, making the replacements
$\alpha\equiv\frac{1}{4\pi}\td{f}{u}$ and $\kappa\equiv\td{M}{u}$, to
arrive at our final expression relating the current and field:
\beq \label{jfinal}
\bj = \alpha(u)\bB+\pom\rho\kappa(u)\be_\phi
\eeq
for a mixed poloidal and toroidal field in axisymmetry.

The two functions $\alpha(u)$ and $\kappa(u)$ govern different aspects
of the magnetic field: firstly, since $\lor=\bj\times\bB$ we have
$\lor=\pom\rho\kappa\be_\phi\times\bB$ (from equation
\eqref{jfinal}) --- i.e., the
Lorentz force is dependent on $\kappa$, and so $\kappa$ governs the
relative contributions of the magnetic and centrifugal forces to the
overall distortion of the star. The role of $\alpha$ is less
clear. From equation \eqref{jfinal} we see
that $\alpha=0$ gives a purely toroidal current and hence poloidal
field, whilst increasing $\alpha$ increases the size of the mixed
toroidal-poloidal term $\alpha{\bf{B}}$ (and so indirectly increases
the toroidal component of the field). However, there is no limit in which the
field is purely toroidal in this formalism. We can thus only expect
$\alpha$ to have some indirect connection with the relative strengths
of the poloidal and toroidal components of the magnetic field.

Following \citet{tomieri}, we choose the functional forms of
$\alpha(u)$ and $\kappa(u)$ as:
\beq\label{kappachoice}
\kappa(u) = \kappa_0=const.,
\eeq
\beq
\alpha(u) = \begin{cases}
               a(u-u_{max})^\zeta & \textrm{if $u>u_{max}$}\\
                        0         & \textrm{if $u\leq u_{max}$,}
            \end{cases}
\label{alphachoice}
\eeq
where $\alpha$ is chosen to ensure there is no exterior current,
$\zeta$ is some constant and $u_{max}$ is the maximum surface value
attained by the streamfunction $u$. Next we combine the definitions
$\alpha\equiv\td{f}{u}$ and $f(u)\equiv \pom B_\phi$ to see that
\beq
\int^u \alpha(u')\ du' = \pom B_\phi
\eeq
--- i.e., we must enforce the continuity of $\int\alpha(u)\ du$ to
ensure the continuity of $B_\phi$. We therefore
choose the lower limit of the integral of $\alpha$ so that
\beq\label{intalphachoice}
f(u) \equiv \int^u \alpha(u')\ du' =
            \begin{cases}
\frac{a}{\zeta+1}(u-u_{max})^{\zeta+1} & \textrm{if $u>u_{max}$,}\\
                        0              & \textrm{if $u\leq u_{max}$.}
            \end{cases}
\eeq
For our chosen functional forms of $\alpha(u)$ and $\kappa(u)$ we see
that for a specific solution we need to choose three constants:
$\zeta$, $a$ and $\kappa_0$. We will later drop the zero subscript,
with the understanding that $\kappa$ always refers to a constant unless
otherwise stated. \citet{tomieri} set $\zeta=1$, but we have found
that a smaller value of $\zeta$ allows for a slightly stronger
toroidal-field component; accordingly, we set $\zeta=0.1$ throughout
this paper, except in comparing our results with previous work
(subsection \ref{TEcompare}).

For the purposes of numerics we seek integral equations; the integral
form of \eqref{newton2} is
\beq \label{enthalpyeq}
H = C - \Phi_g + \Phi_r + \int_0^{\pom\Aph}\kappa(u')\ du',
\eeq
where $C$ is an integration constant and
\beq
H(\br)=\int_0^{P(\br)} \frac{dP'}{\rho(P')}
\eeq
is the enthalpy.

The integral form of Poisson's equation \eqref{diffpoisson} is:
\beq \label{intpoisson}
\Phi_g(\br) = -G \int
                \frac{\rho(\br')}{|\br-\br'|}\ d\br'.
\eeq

Finally, using the current relation \eqref{jfinal} an integral
equation for the magnetic field may be found (see \citet{tomieri} for
details):
\beq \label{finalAph}
\Aph(\br)\sin\phi
 = \int
     \frac{4\pi\frac{\alpha(\pom'\Aph')}{\pom'}\int_0^{\pom'\Aph'}\alpha(u)\ du
           +\kappa\rho \pom'} 
           {|\br-\br'|}
   \sin\phi'\ d\br'.
\eeq

With the three equations \eqref{enthalpyeq}, \eqref{intpoisson} and
\eqref{finalAph} it is possible to calculate stationary configurations
of magnetised rotating stars (together with the specified constants
$a$ and $\kappa$).

\subsection{Toroidal-field formalism}

For a purely toroidal field we have ${\bf{B}}=B_\phi
{\bf{e}}_\phi$. In this case, the $\GSu$ term disappears from equation
\eqref{jandBgen} and the Lorentz force reduces to the form
$\lor=B_\phi {\bf j}_{pol}\times {\bf e}_\phi$. Comparing these two
expressions, one can show that $B_\phi$ is related to
$\gamma\equiv\rho\pom^2$ through some arbitrary function $h$ (which
must vanish outside the star):
\beq
B_\phi = \frac{1}{\pom} h(\gamma)
\eeq
--- see section \ref{tor_formal} for details. The magnetic potential
for this case is:
\beq
M = -\frac{1}{4\pi}\int_0^{\rho\pom^2}
                      \frac{h}{\gamma}\td{h}{\gamma}\ d\gamma,
\eeq
where $\nabla M=\lor/\rho$ as before.

For simplicity we choose $h(\rho\pom^2)=\lambda\rho\pom^2$ where
$\lambda$ is a constant we specify for each code run. With this
choice of $h$ we then have $B_\phi=\lambda\rho\pom$ and
$M=-\lambda^2\rho\pom^2$, so that the first integral of the Euler
equation becomes
\beq
H = C-\Phi+\textstyle{\frac{1}{2}}\Omega^2\pom^2
          -\frac{1}{4\pi}\lambda^2\rho\pom^2.
\label{toreuler}
\eeq
This equation, together with Poisson's equation, is sufficient to find
numerical solutions; the toroidal-field case is thus simpler than the
mixed-field formalism, which also had an extra equation for the
magnetic field.

\subsection{Restrictions on the magnetic functions}
\label{mag-restrict}

In the appendix (outlined in the above sections) we show that for
axisymmetric perfect MHD in a
fluid, the equations reduce to a mixed-field case (with two magnetic
functions $\alpha(u)$ and $\kappa(u)$) and a purely 
toroidal case (with a magnetic function $h(\gamma)$). Although the
magnetic functions appear to be arbitrary, there are a number of
restrictions on their functional forms, on either physical
grounds or because they result in trivial solutions.

The functions $\alpha(u)$ and $h(\gamma)$ (where $u=\pom\Aph$ and
$\gamma=\rho\pom^2$ as before) govern the toroidal fields
in the two cases, and so both must necessarily vanish outside the star
to avoid having exterior currents. Since the streamfunction $u$ does
not vanish at the star's surface, we follow \citet{tomieri} in
defining $u_{max}$ to be the
maximum surface value of $u$ and then choose $\alpha(u)$ to be a power
of $(u-u_{max})$, which \emph{does} vanish outside the star. There
does not appear to be any other functional form for $\alpha$ which
vanishes outside the star and is dependent only on $u$, so we conclude
that \eqref{alphachoice} is the only acceptable choice for
$\alpha(u)$. The functional form of $h$, similarly,
appears restricted. To vanish outside the star $h(\gamma)$ cannot
contain a constant piece, so let us consider a functional form of
$h(\gamma)=\lambda\gamma^\chi$ where $\lambda$ and $\chi$ are
constants. However, if $\chi<\frac{1}{2}$ then
$B_\phi=\lambda\gamma^\chi\pom^{-1}=\lambda\rho^\chi\pom^{2\chi-1}$ will
diverge at the origin, so we discard these choices. Additionally, we
find that if $\chi>1$ is chosen, then the field iterates to zero
in our numerical scheme, leading us to choose $h(\gamma)=\lambda\gamma$.

Finally, the function $\kappa(u)$ is theoretically allowed to depend
on the streamfunction $u$, but if it is chosen as anything other than
a constant then, as for $h(\gamma)$, we find that the configuration
iterates to a zero-field solution. This may be a limitation of our
numerical scheme rather than a physical restriction, but in either
case our HSCF-scheme solutions are limited to those with $\kappa$
being equal to some constant.

We conclude from this that, in fact, the choices made for our
functional forms are not specialised ones and (at least within our
scheme) do not result in the
exclusion of physically valid solutions. Rather, we believe that our
results are quite generic to perfectly conducting polytropes in
axisymmetry.

\section{Numerics and calculating various quantities}

\subsection{Numerical scheme}

Our code uses the Hachisu self-consistent field (HSCF)
method (\citet{hachisu}, extended to magnetised configurations
by \citet{tomieri}) to
iteratively find a stationary solution to the hydromagnetic
equilibrium equation \eqref{newton2}. Specifically, one specifies
a polytropic index $N$, magnetic functions $\alpha(u)$ and $\kappa(u)$
and the ratio of polar to equatorial radii $r_p/r_{eq}$, and the code
determines the angular velocity, density distribution and other
quantities consistent with the user's input parameters.

The iterative steps for our extended (mixed-field) HSCF scheme are:\\
 \\
1. Make an initial guess of $\rho$=const.;\\
2. Find $\Phi_g$ from Poisson's equation \eqref{intpoisson};\\
3. Guess $\Aph$=const.;\\
4. Find an improved form of $\Aph$ from equation \eqref{finalAph} and
   the earlier guesses for $\rho$ and $\Aph$ (this is the iterative
   step for $\Aph$);\\
5. Find $\Omega_0^2$ and $C$ from the boundary condition that the
   enthalpy must vanish at the surface of the star; this requires
   the potentials $\Phi_g$ and $\Aph$ found earlier and a
   user-specified axis ratio $r_p/r_{eq}$;\\
6. We now know all right-hand side terms in \eqref{enthalpyeq}; use
   the equation to determine the enthalpy at all points in the star;\\
7. Find the new (improved) estimate for the density distribution
   using $\rho_{new}(\br) = \brac{\frac{H(\br)}{H_{max}}}^N$ where
   $N$ is the polytropic index and $H_{max}$ the maximum value of
   enthalpy attained in the star;\\
8. As the iterative step, return to step 1 but use $\rho=\rho_{new}$
   instead of the earlier density distribution ($\rho$=const. for the
   first cycle). At step 3 in the new cycle, use the
   `new' form of $\Aph$ calculated in step 4 of the previous cycle.\\
 \\
This sequence of steps is repeated until the code has achieved
satisfactory convergence in $H_{max}$, $\Omega^2_0$ and $C$. The
toroidal-field scheme is similar to the one above except that
the magnetic field is directly related to the density by
$B_\phi=\lambda\rho\pom$ for pure toroidal fields. For this reason
there is no separate iteration in the magnetic field and steps 3 and 4
are no longer needed. The magnetic field only enters in step 6, where
the enthalpy $H$ is found from the pure-toroidal equation
\eqref{toreuler} instead of the mixed-field version
\eqref{enthalpyeq}.

\subsection{Magnetic energy and field strength}

We will wish to calculate the magnetic energy $\emag$ of the star in the
code, to compare different configurations and also to calculate a virial
test (see section \ref{virial} and figure \ref{convtests}). The
familiar definition of $\emag$ is
\beq
\emag = \int\limits_{\textrm{all space}} \frac{B^2}{8\pi} \ \ d\br,
\eeq
but this is not suited to numerical evaluation, since the integrand
only decays at infinite distance; our numerical integration is over a
finite radius and so this definition would introduce truncation error.
However a physically equivalent definition for $\emag$, more useful
here, is
\beq
\emag = \int\limits_{\textrm{all space}} \br\cdot\lor \ \ d\br
\eeq
--- since $\lor$ has compact support (through its $\rho$ dependence)
the above integrand will vanish outside the star.

For a measure of the magnetic field strength of the star, we define a
volume-averaged magnetic field $\Bav$ through
\beq \label{Bphys}
\Bav^2 \equiv \frac{1}{V}\!\!\int\limits_{\textrm{all space}}\!\! B^2\ d\br
          =   \frac{8\pi\emag}{V}.
\eeq

\subsection{Virial test}
\label{virial}

\begin{figure}
\begin{center}
\begin{minipage}[c]{0.75\linewidth}
\psfrag{a}{(a)}
\psfrag{b}{(b)}
\psfrag{c}{(c)}
\psfrag{MP}{$MP$}
\psfrag{VC}{$VC$}
\includegraphics[width=\linewidth]{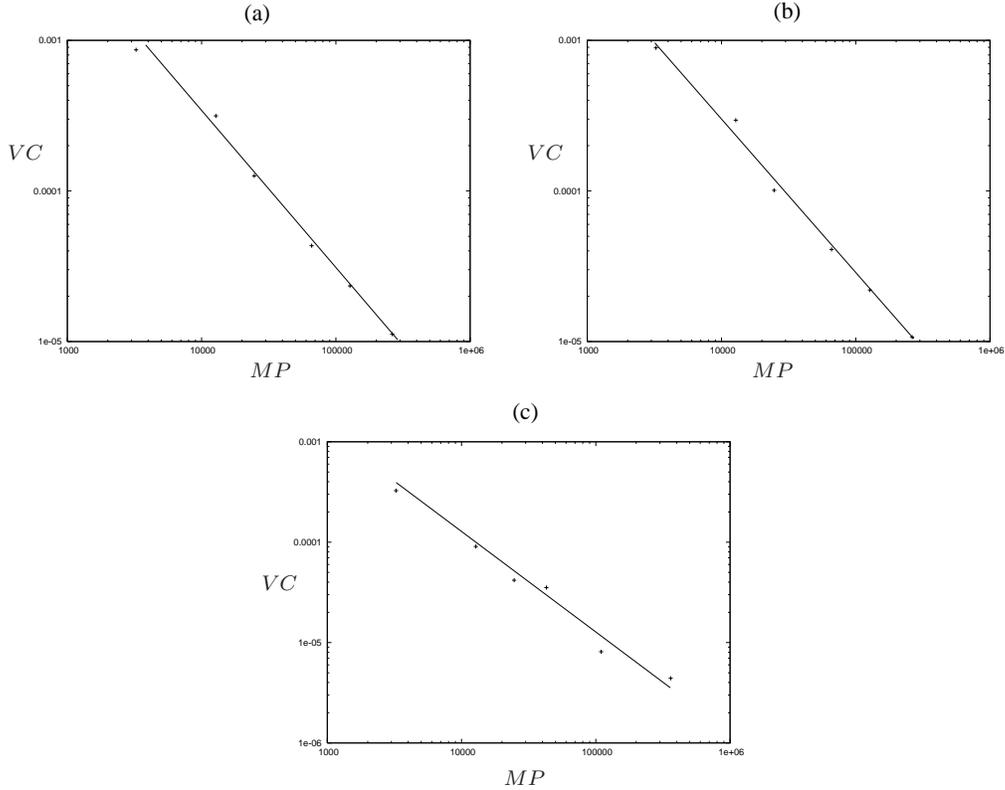}
\end{minipage}
\caption{\label{convtests}
         Convergence tests for: (a) a purely poloidal field, no
         rotation, axis ratio of 0.2; (b) mixed field with 2\%
         toroidal field, no rotation, axis ratio of 0.2; (c) purely
         toroidal field, no rotation, axis ratio of 1.05. Here $VC$
         is the virial test result and $MP$ the number of mesh
         points. Since $VC$ decreases as $MP$ increases, we see that
         the code is convergent.}
\end{center}
\end{figure}

We may use the scalar virial theorem (see, e.g., \citet{shapteuk}) as
a test of convergence for the code. For a rotating magnetised
self-gravitating fluid the virial theorem states that
\beq
\frac{1}{2}\td{^2 I}{t^2} = 2T+\emag+\frac{3U}{N}+W
\eeq
where $I$ is the moment of inertia about the rotation axis and
$T,\emag,U$ and $W$ are the kinetic,
magnetic, internal and gravitational energies, respectively. For our
stationary star $I$ has no time variation so the first term is
zero. Given this, we expect the
various energies for our star to satisfy
\beq
2T+\emag+\frac{3U}{N}+W = 0.
\eeq
Calculating the quantity on the left-hand side of the above equation
tells us the absolute deviation from zero, but we need to know the
relative error. A value of $2T+\emag+3U/N+W = 10^{-5}$ would appear to
indicate acceptable accuracy, but if the individual energies are of
order $10^{-4}$ then the relative error is unacceptable: around
10\%. For this reason we normalise by dividing through by $W$ and
define our virial test result $VC$ as
\beq \label{virialtest}
VC \equiv \frac{|2T+\emag+3U/N+W|}{|W|}
\eeq
--- the smaller the value of $VC$, the greater the code's accuracy. We
use $VC$ in our convergence testing, figure \ref{convtests}. In the
figure we see that as grid resolution is increased the virial test
result decreases; in particular, since the gradient of each plot is
approximately $-1$ we conclude that the code is first-order convergent.

\subsection{Comparison with previous work}
\label{TEcompare}

As a confirmation of our results, we compare with table 4
from \citet{tomieri}. Their results are nondimensionalised by dividing
by appropriate powers of $\rho_{max}$, $r_{eq}$ and $4\pi G$ and these
dimensionless quantities are denoted by a hat; for example
\beq
\hat{\Omega}^2=\frac{\Omega^2}{4\pi G \rho_{max}}.
\eeq
For comparison with their results we must also use $\zeta=1$ instead
of $\zeta=0.1$ as the exponent in the functional form
of $\alpha$ from \eqref{alphachoice}. Taking this into
account we find that for a $N=1.5$ polytrope, with $\hat\kappa=0.4$ and
$\hat{a}=200$, we have the sequence of configurations given in table
\ref{comparetable}.

\begin{table}
\begin{center}
\caption{\label{comparetable}
         Dimensionless quantities for a sequence of stars with
         $N=1.5,\hat\kappa=0.4,\hat{a}=200$ and $\zeta=1$.}
\begin{tabular}{cccccccc}
$r_p/r_{eq}$ & $\emag/|W|$ & $U/|W|$ & $T/|W|$ & $|\hat{W}|$ &
$\hat\Omega^2$ & $\hat{M}$ & $VC$\\
\hline
0.588 & 0.144 & 0.284 & 1.21e-03 & 4.81e-02 & 5.14e-04 & 0.831 & 2.97e-05\\
0.55  & 0.151 & 0.276 & 1.11e-02 & 4.59e-02 & 4.53e-03 & 0.811 & 3.10e-05\\
0.50  & 0.165 & 0.264 & 2.11e-02 & 4.32e-02 & 8.01e-03 & 0.787 & 3.33e-05\\
0.45  & 0.189 & 0.255 & 2.27e-02 & 4.01e-02 & 7.72e-03 & 0.763 & 3.63e-05\\
0.40  & 0.222 & 0.252 & 1.19e-02 & 3.58e-02 & 3.45e-03 & 0.729 & 4.02e-05\\
0.371 & 0.242 & 0.252 & 1.10e-03 & 3.31e-02 & 2.89e-04 & 0.705 & 4.32e-05
\end{tabular}\\
\end{center}
\end{table}

Our highest and lowest axis ratios ($0.588$ and $0.371$) differ
slightly from those of \citet{tomieri} (who have $0.589$ and $0.372$),
so we cannot make a direct comparison for these values. However for
the other four axis ratios our values agree to within $\sim 8\%$ for
$\hat\Omega^2$ and $T/|W|$, and to within $\sim 1\%$ in all other
quantities. We also show our results for the virial test, showing that
all our results have relative errors of $\sim 10^{-5}$.

\subsection{Toroidal and poloidal energies for the mixed case}

The code variables $\kappa$ and $\alpha$ are related to the ratio of
toroidal to poloidal field strength, but in a very nontrivial
manner. To get a more intuitive, physical, measure of their respective
strengths we would like to know the part of the magnetic energy
contained in the poloidal and toroidal fields, $\epol$ and $\etor$,
respectively.

Since the total magnetic energy is given by
\beq
\emag = \frac{1}{8\pi} \int {\bf{B\cdot B}}\ d\br
      = \frac{1}{8\pi} \int \brac{B_\pom^2+B_\phi^2+B_z^2} \ d\br,
\eeq
we define the poloidal energy by
\beq
\epol = \frac{1}{8\pi}\int\brac{B_\pom^2+B_z^2} \ d\br
      = \frac{1}{2}\int_0^1\int_0^\infty\brac{B_\pom^2+B_z^2}\ drd\mu
\eeq
and the toroidal energy by
\beq
\etor = \frac{1}{8\pi} \int B_\phi^2 \ d\br
      = \frac{1}{2} \int_0^1\int_0^\infty B_\phi^2 \ drd\mu
\eeq
where the integration here is over spherical polars $r$ and
$\mu\equiv\cos\theta$. Note that since $B_\phi=0$ outside the star,
the toroidal-energy integral only needs to be evaluated over the
stellar interior. For our mixed-field configurations we use
$\etor/\emag$ as a measure of the proportion of toroidal field.

\subsection{Ellipticity}

For the code we specify the axis ratio $r_p/r_{eq}$, which
measures the distortion of the star's surface. For a measure of the
distortion of the whole mass distribution of the star we define an
ellipticity $\eps$ through the (unreduced) quadrupole
moments at the equator $I_{eq}$ and the poles $I_p$:
\beq \label{ellip}
\eps \equiv \frac{I_{eq}-I_p}{I_{eq}}.
\eeq

\subsection{Constructing physical sequences of stars}
\label{redim}

To make a meaningful study of a group of different equilibrium
configurations requires ensuring that we are always comparing the
effects of magnetic fields and rotation in the same physical star: we
do this by
ensuring that we work with sequences of constant (physical) mass and
the same equation of state --- i.e. both the same polytropic index $N$
\emph{and} polytropic constant $k$. We note that other intuitively
sensible choices, for example just fixing the equatorial radius or central
density, would mean comparing stars of either different mass or
different equation of state. This would make quantifying the effects of
magnetic fields and rotation more difficult.

We fix our neutron star mass to the generic value of
$M=1.4M_\odot=2.8\times 10^{33}$ g. For the equation of state we work
with $N=1$ polytropes and fix $k$ by requiring that the radius $R$ of the
equivalent unmagnetised nonrotating (and hence spherical) star is 10
km. We will term this star the `background' star, with the
understanding that this refers to a configuration without magnetic
fields or rotation, rather than having any perturbation theory
connotations. Using the ($N=1$) polytropic relation
$R=\sqrt{\pi k/2G}$ we see that this
gives a polytropic constant of $4.25\times 10^4$
g${}^{-1}$cm${}^5$s${}^{-2}$. For the rest of this paper, when
we quote physical parameters they will be for our `canonical neutron
star' with $M=1.4M_\odot$ and $k=4.25\times 10^4$
g${}^{-1}$cm${}^5$s${}^{-2}$.

For the plots where we have used different polytropic indices the
redimensionalising is less straightforward, as the relation between
spherical radius $R$ and $k$ also includes powers of the
`background' central density $\rho_c$ (see \citet{stell-struct} for the
required polytropic relations). For these cases we again fix the
mass at $1.4M_\odot$ and fix the background central density at
$\rho_c=2.19\times 10^{15}$ g cm${}^{-3}$ --- the same value as for
the background $N=1$ star discussed above. This then fixes $R$ and
$k$.

\section{Magnetic field configurations}
\label{magfields}

\begin{figure}
\begin{center}
\begin{minipage}[c]{0.7\linewidth}
\psfrag{a}{(a)}
\psfrag{b}{(b)}
\psfrag{c}{(c)}
\psfrag{d}{(d)}
\includegraphics[width=\linewidth]{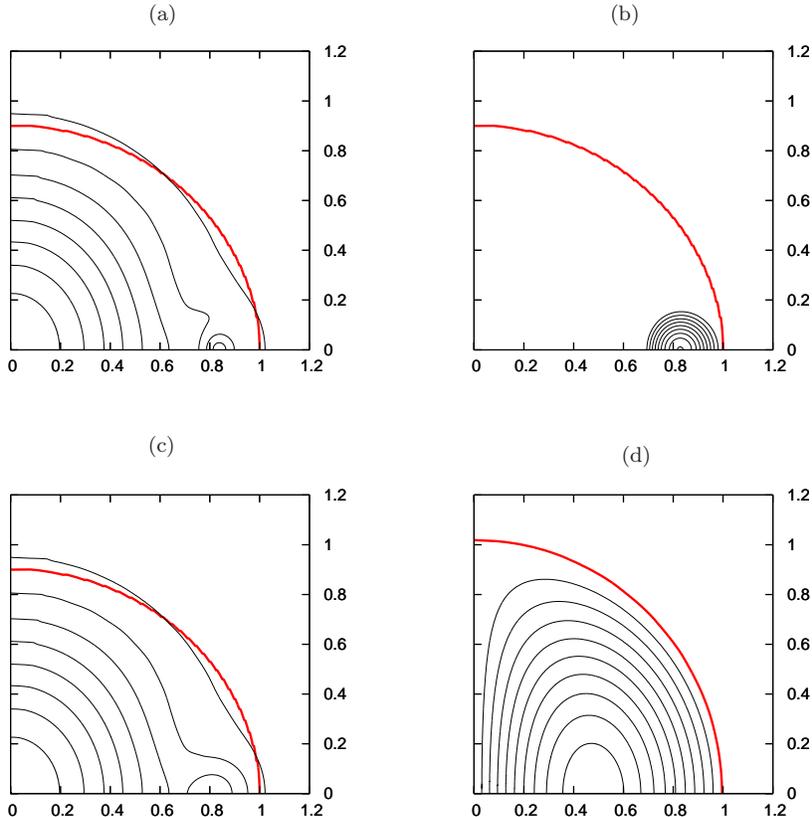}
\end{minipage}
\caption{\label{polandtorcomps}
         Contours of magnetic field strength. Plots (a), (b), (c) are
         (respectively) the poloidal, toroidal and total magnetic
         field strength for our canonical NS with a mixed-field
         configuration consisting of 3.0\%
         toroidal field. Plot (d) is for magnetic field strength in a
         purely toroidal-field star. Note how the toroidal field in
         this case is much more extensive than in the mixed-field plot
         (b). In plots (a) and (c) the maximum field strength is
         $5.5\times 10^{17}$ G (at the origin) and the contour
         separation is $5.5\times 10^{16}$ G. For plots (b) and (d)
         the maximum field occurs in the centre of the torus bounding
         the toroidal field; the
         maximum values are $2.6\times 10^{17}$ G and $2.8\times
         10^{17}$ G, with contour separations of $2.9\times 10^{16}$ G
         for both plots. The bold red line in
         each plot represents the star's surface; the values on the
         axes show the nondimensional radius $r/r_{eq}$ (where $r$ is
         the physical radius).}
\end{center}
\end{figure}

\begin{figure}
\begin{minipage}[c]{\linewidth}
\begin{center}
\includegraphics[width=0.4\linewidth]{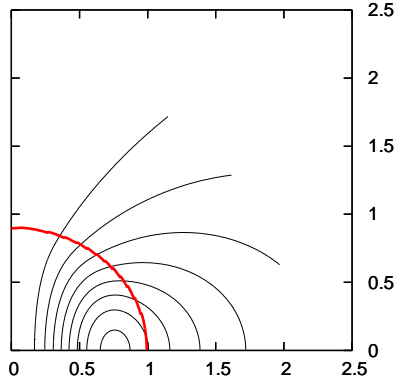}
\end{center}
\end{minipage}
\caption{\label{ucontour}
         Contours of the streamfunction $u$ for a purely
         poloidal-field star; these contours are parallel to magnetic
         field lines and so represent the direction of the field. The
         surface of the star is the bold red line. Field lines for the
         toroidal component of a mixed-field star, or for purely
         toroidal stars, would go into the page and hence are
         not plotted here.}
\end{figure}

\begin{figure}
\begin{minipage}[c]{\linewidth}
\begin{center}
\psfrag{N}{$N$}
\psfrag{B_ratio}{$\displaystyle\frac{B_p}{\Bav}$}
\includegraphics[width=0.4\linewidth]{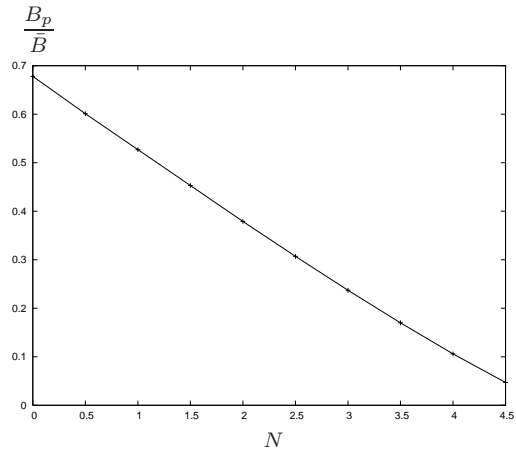}
\end{center}
\end{minipage}
\caption{\label{N-Bratio}
         Left: the ratio of polar field to volume-averaged field,
         $B_p/\Bav$, as a function of the polytropic index
         $N$. The plot is for purely poloidal fields in non-rotating
         stars, all with an axis ratio of $0.996$. Note that if the
         field was purely toroidal then this ratio would be zero,
         regardless of $N$.}
\end{figure}

With the formalism described above, we are able to examine the field
configurations generated in axisymmetric perfectly conducting
polytropes. Since neutron star matter is thought to have high
conductivity and be roughly approximated by an $N=1$ polytrope, the
field structures shown here should have some similarity to those in
real neutron stars --- although the field strengths here are
considerably higher than those that have been observed so far. The
plots in this section show contours of the magnetic field strength
given by $|\bB|=\sqrt{\bB\cdot\bB}$, and of the poloidal and
toroidal components, $|{\bf B}_{pol}|=\sqrt{B_{\pom}^2+B_z^2}$ and
$|{\bf B}_{tor}|=|B_\phi|$. All of the magnetic-field results
presented here (and discussed in this section) are for non-rotating $N=1$
polytropes, unless otherwise stated. In addition, we have concentrated
on mixed-field configurations here, since there are strong indications
from both theory
\citep{taylerpol,taylertor,wright} and simulations \citep{braithwaite}
that both purely poloidal and purely toroidal fields are
generically unstable.

In figure \ref{polandtorcomps} we plot the poloidal (plot (a)) and
toroidal (plot (b))
components of a mixed-field star and the total field of the
configuration (plot (c)). We also
plot the field structure of a star generated from our purely
toroidal-field formalism for comparison (plot (d)). We see that the
poloidal field pervades most of the interior of the mixed-field star,
as well as extending outside it. This component of the 
field is highest in the centre and only goes to zero in a small region
at the edge of the star (seen as the pair of semicircular contours on
the equator at $x\sim 0.8$); \cite{taylerpol} call this zero-field
point the `magnetic axis'. By contrast the toroidal field is wholly
contained within
this small region where the poloidal field vanishes; this region is
dictated by the functional form of $\alpha(u)$ that we use. All
configurations shown in this section are non-rotating, but rotation
does not greatly affect the nature of the magnetic field.

Comparing plots (b) and (d) in figure \ref{polandtorcomps} we see
that, although the maximum field strengths and contours are of similar
magnitude in the two cases, the field in the pure-toroidal case
extends over a far larger
region of the star than the toroidal part of the mixed-field
configuration.

Figure \ref{polandtorcomps} shows the
\emph{magnitude} of the magnetic field at a particular
point; in figure \ref{ucontour} we show the direction of a typical
poloidal field by
plotting contours of the streamfunction $u$. These contours are
parallel to magnetic field lines, from section \ref{mix_formal} of the
appendix. Since a purely toroidal field has direction vector
$\be_\phi$, the field lines would go into the page in the $x-z$ plane
we employ here (they would form concentric circles in the $x-y$
plane). Mixed-field lines lie in neither plane so we have not shown
them here.

Lastly in this section, figure \ref{N-Bratio} shows the dependence of
the ratio $B_p/\Bav$ on the polytropic index $N$; we find that
there is an approximately linear relationship between the two, and 
for all polytropic indices $B_p/\Bav$ is of the same order
of magnitude. For $N=1$, $B_p/\Bav\approx 0.5$, suggesting that
neutron stars (approximated as $N=1$ polytropes) with purely poloidal
fields are likely to have a $\Bav$ around double the polar field $B_p$.

\subsection{The relationship between $a$ and $\etor/\emag$}

\begin{table}
\begin{center}
\caption{\label{tortable} Comparing parameters related to the influence
of the toroidal component in a mixed-field star with axis ratio $0.9$.}
\begin{tabular}{ccccc}
$a$ & $\etor/\emag$ & $\emag/|W|$ & $\eps$ & $B_p/\Bav$\\
\hline
0  & 0.00     & 2.43e-02 & 0.216 & 0.580 \\
10 & 9.87e-03 & 2.55e-02 & 0.216 & 0.554 \\
20 & 3.02e-02 & 2.82e-02 & 0.213 & 0.504 \\
30 & 3.96e-02 & 2.93e-02 & 0.204 & 0.484 \\
40 & 4.05e-02 & 2.92e-02 & 0.196 & 0.488 \\
50 & 3.86e-02 & 2.88e-02 & 0.191 & 0.495 \\
\end{tabular}\\
\end{center}
\end{table}

As mentioned earlier, we can increase the proportion of toroidal field
in the mixed-field configurations only indirectly, by varying the code
parameter $a$ from equation \eqref{alphachoice}. In table
\ref{tortable} we show the effect of changing this
parameter, for a non-rotating star with axis ratio
$r_p/r_{eq}=0.9$. One would expect that increasing $a$ would
increase the toroidal portion of the field, which in turn would lead
to a decrease in oblateness (since toroidal fields induce prolate
distortions); one would also expect a reduction in the ratio
$B_p/\Bav$ (since more of the field 
is toroidal and hence does not extend outside the star). Looking at
the table, we see all of these effects do occur as the value of $a$ is
increased, up until the $a=40$ configuration. At this point the larger
value of $a$ is no longer reflected in stronger toroidal-field
effects. In all cases changing $a$ does not strongly affect the value
of $\emag/|W|$, confirming our expectation that it is the variation in
the toroidal component which affects ellipticity and $B_p/\Bav$,
rather than simply a reduction in $\emag/|W|$. Finally, we note that
even for the highest values of $a$, the relative contribution of the
toroidal portion of the field is very small --- only 4\% of the total
here. We shall see later that this is a generic feature of our
formalism together with our boundary condition, where poloidal fields
extend outside the star but toroidal ones vanish at the surface.

\section{Magnetic and rotational distortions}

\begin{figure}
\begin{minipage}[c]{\linewidth}
\begin{center}
\psfrag{(a)}{(a)}
\psfrag{(b)}{(b)}
\psfrag{(c)}{(c)}
\psfrag{(d)}{(d)}
\includegraphics[width=0.7\linewidth]{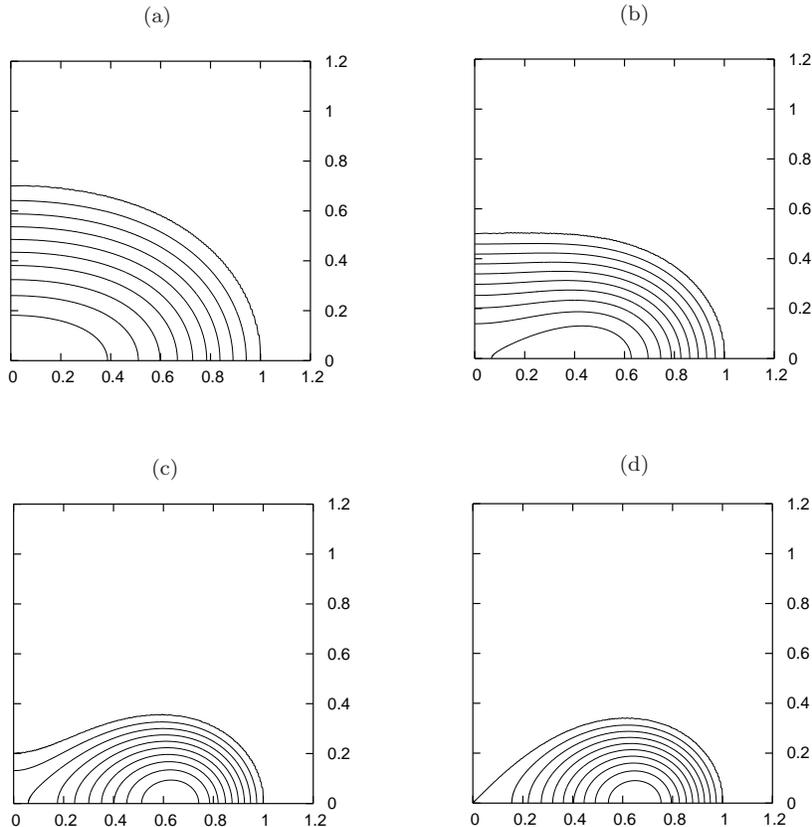}
\end{center}
\end{minipage}
\caption{\label{torus-shape}
         Nonrotating $N=1$ polytropes distorted by the effect of a poloidal
         magnetic field, with axis ratios of $0.7,0.5,0.2,0.0$ (from
         (a) to (d)). For increasing distortion the maximum density
         moves away from the centre and the density distribution
         becomes toroidal (in the sense that the maximum density moves
         away from the centre of the star). As before, the numbers on
         the axes are dimensionless, but for our canonical NS
         $r_{eq}=11.4,12.9,16.2,17.0$ km for plots (a)-(d).}
\end{figure}

\begin{figure}
\begin{minipage}[c]{\linewidth}
\begin{center}
\psfrag{(a)}{(a)}
\psfrag{(b)}{(b)}
\psfrag{(c)}{(c)}
\psfrag{(d)}{(d)}
\includegraphics[width=0.7\linewidth]{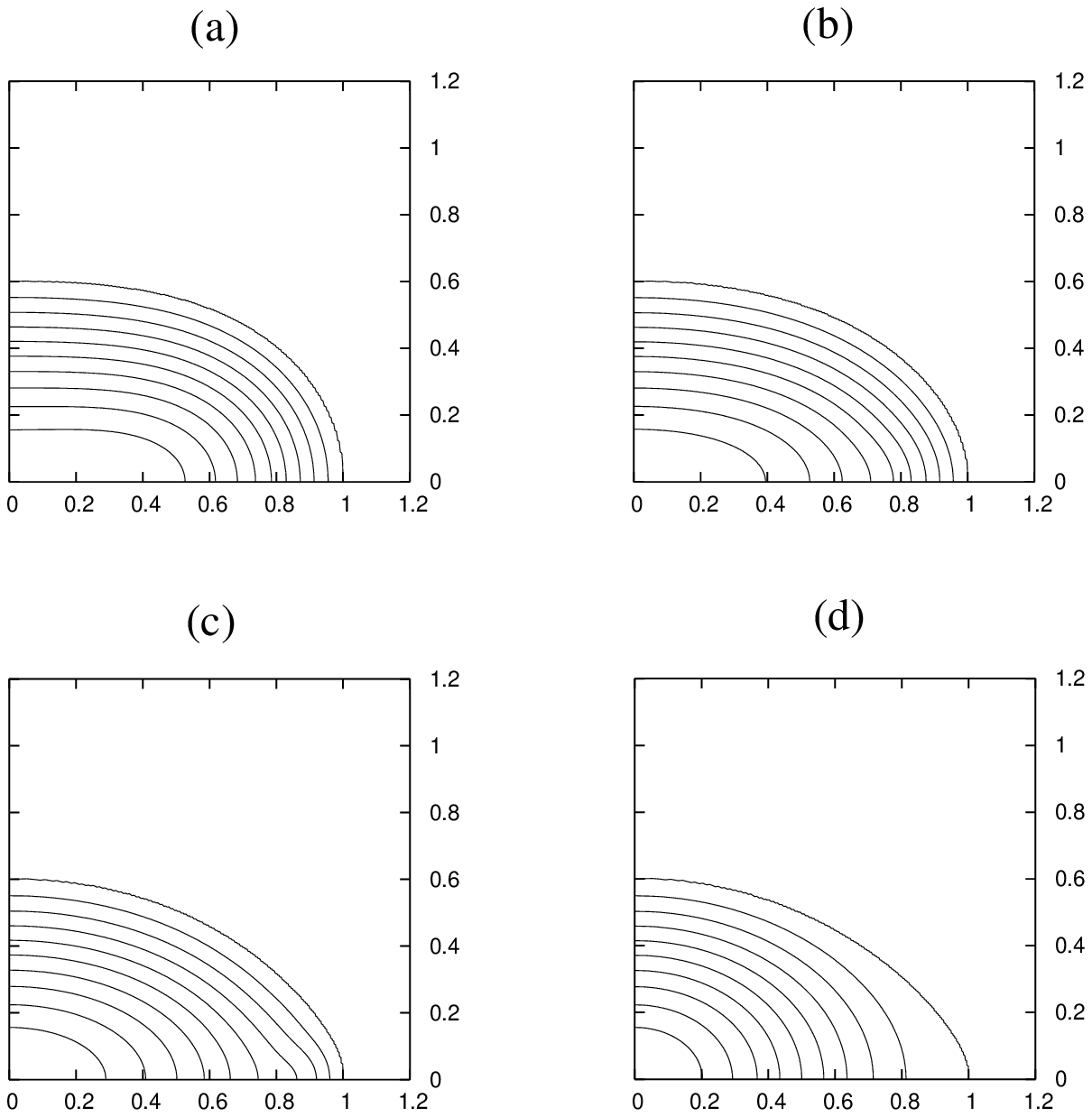}
\end{center}
\end{minipage}
\caption{\label{r06dens}
         Density contours in an $N=1$ polytropic star with axis ratio
         of 0.6, with different sources of distortion. Plots (a),
         (b) and (c) are nonrotating configurations with,
         respectively: purely poloidal field, mixed-field with 3.4\%
         toroidal field, mixed-field with 5.5\% toroidal field. Plot
         (d) is for a purely rotationally-distorted star with no
         magnetic field. All stars have the canonical mass of $1.4M_\odot$,
         with equatorial radii of $12.1,12.5,13.2,14.4$ km for stars
         (a)-(d), respectively. We note 
         that whilst a purely poloidal field tends to push the maximum
         density away from the centre, both toroidal field components
         and rotation have the effect of increasing the equatorial
         radius and making the star more diffuse.}
\end{figure}

Having looked at field configurations, we now turn to the distortions
these fields produce in the star's mass distribution. For purely
poloidal fields we confirm previous work that these fields induce an
oblate distortion; the surface shapes of such stars are thus similar
to those of rotationally distorted stars. However, the interior
density distributions are very different: centrifugal forces tend to
leave a smaller high-density central region, whilst the Lorentz
force acts to pull the point of maximum density away from the centre into a
maximum-density ring. In the extreme limit where the ratio $r_p/r_{eq}
\to 0$, the star actually becomes a torus (figure
\ref{torus-shape}). For mixed fields, the effect of increasing
the toroidal component is similar to the effect of adding rotation: it
tends to push the maximum density region back to the centre --- see
figure \ref{r06dens}. Note that both the mixed-field stars shown are
oblate though, due to the dominance of the poloidal component;
stronger toroidal fields tend to make stars prolate, but our formalism
and boundary condition seem to generate mixed fields with weak
toroidal components only (the 5.5\%-toroidal field of figure
\ref{r06dens} plot (c) is relatively strongly toroidal, within this
context).

\begin{figure}
\begin{minipage}[c]{0.48\linewidth}
\begin{center}
\psfrag{distortion}{\emph{distortion}}
\psfrag{eps}{$\eps$}
\psfrag{1-r_B}{$1-\frac{r_p}{r_{eq}}$}
\psfrag{Bsq}{$\displaystyle{\brac{\frac{\Bav}{10^{17}\textrm{ G}}}^2}$}
\includegraphics[width=\linewidth]{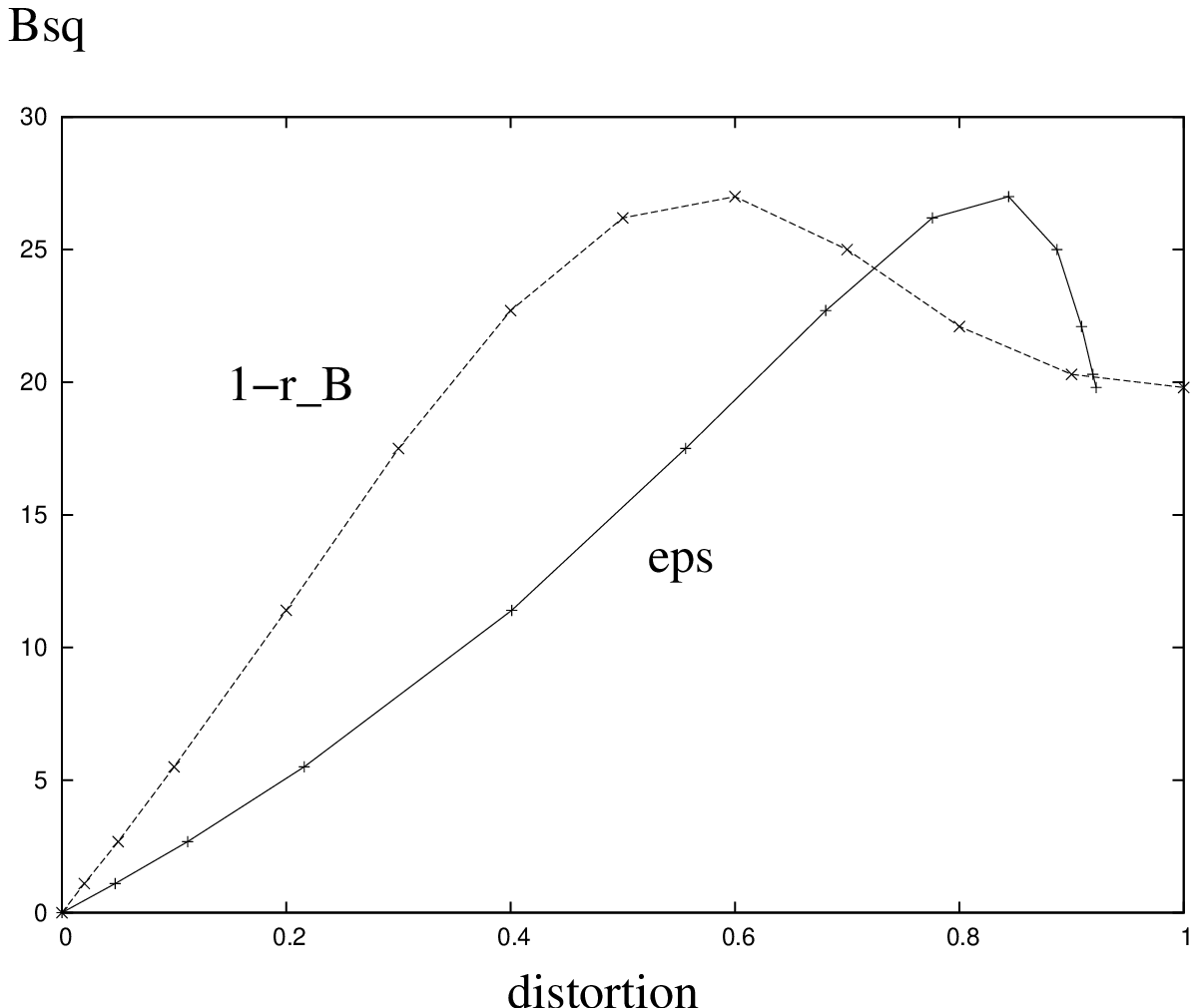}
\end{center}
\end{minipage}
\begin{minipage}[c]{0.48\linewidth}
\begin{center}
\psfrag{eps}{$\displaystyle\eps$}
\psfrag{epsi}{$-\displaystyle\eps$}
\psfrag{B}{$\displaystyle{\brac{\frac{\Bav}{10^{17}\textrm{ G}}}^2}$}
\includegraphics[width=\linewidth]{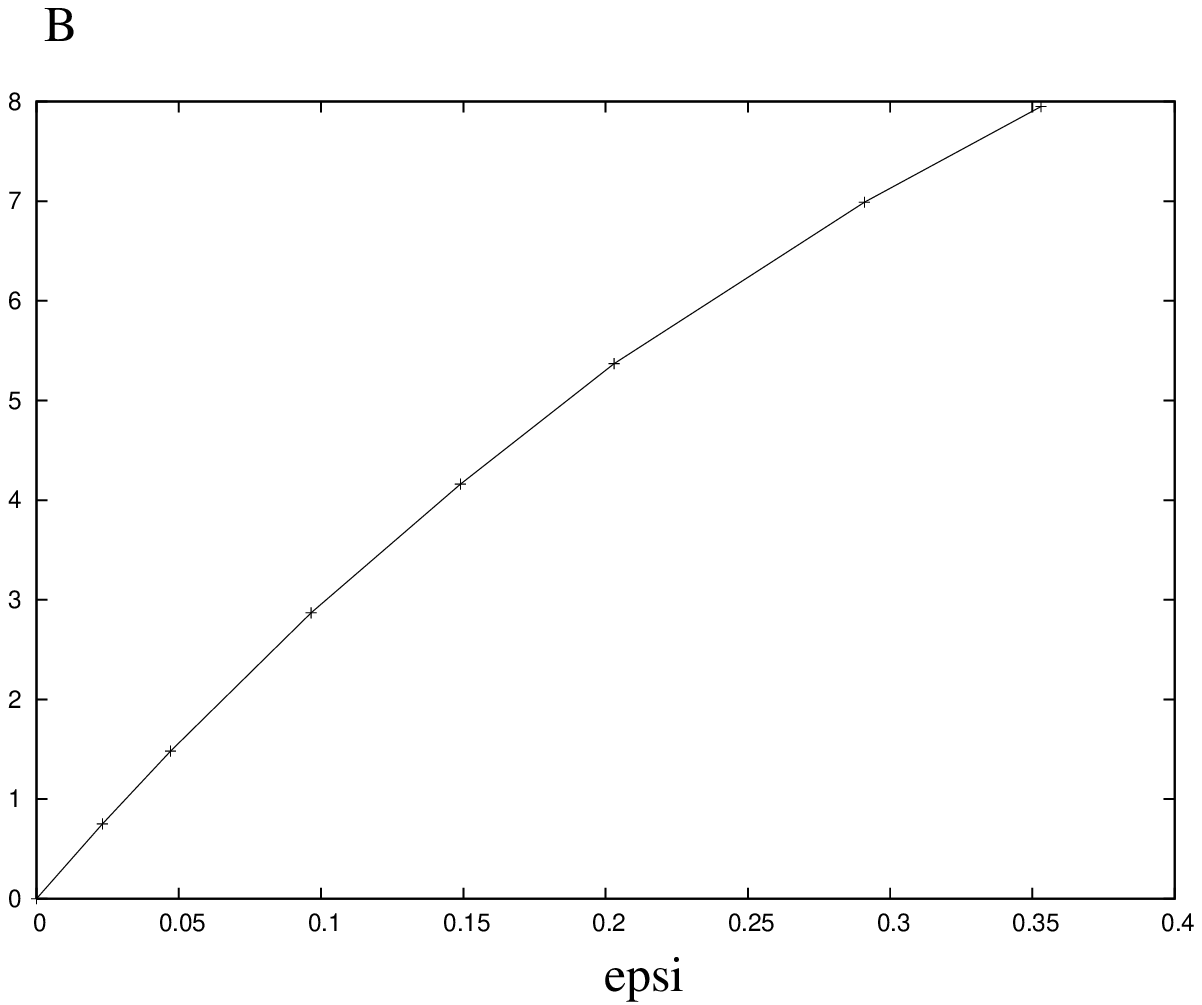}
\end{center}
\end{minipage}
\caption{\label{ellipvsB2}
         Left: a graph showing how (poloidal) magnetic distortions
         vary with the field strength. $1-r_p/r_{eq}$ is the surface
         distortion, whilst $\eps$ represents the distortion of the
         density distribution, as defined in equation
         \eqref{ellip}. Note that the required field strength
         peaks for $1-r_p/r_{eq}\sim 0.6$
         or $\eps\sim 0.8$ and then drops slightly for more extreme
         distortions. For small distortions we see that there is a
         roughly quadratic dependence on the field strength.
         Right: toroidal-field distortions versus $B^2$. In this case
         we only use $\eps$ to gauge the level of distortion, as the
         surface shapes remain nearly spherical.}
\end{figure}

For weak fields and small distortions, perturbation theory results
suggest that the ellipticity of a star should depend linearly on
$B^2$. With our non-linear code we are able to
check this, and see how well the perturbative result holds as field
strengths are increased; this is plotted for both poloidal and
toroidal fields in figure \ref{ellipvsB2}. The results depart slowly
from the linear regime to begin with, but in the poloidal-field case
the field strength required reaches a peak and then decreases again, for
increased ellipticity. This peak seems to correspond to roughly the
point at which the maximum density is pulled out into a ring, making
the star's density distribution toroidally-shaped. We speculate that
for very low axis ratios (i.e. very strong fields), this
toroidally-shaped density is a more stable, lower-energy state than
one where the maximum density remains at the centre.

\begin{figure}
\begin{minipage}[c]{\linewidth}
\begin{center}
\includegraphics[width=0.7\linewidth]{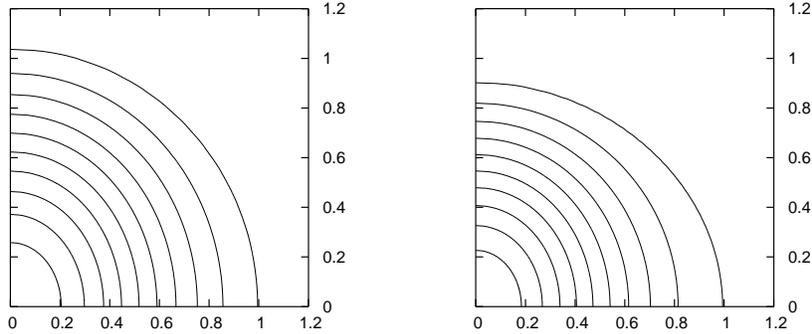}
\end{center}
\end{minipage}
\caption{\label{tordens}
         Two stars with toroidal magnetic fields. The left-hand
         configuration is a non-rotating star (and hence has a prolate
         density distribution), whilst the right-hand one is the same
         physical star but with rotation added, with an oblate surface
         shape but an overall ellipticity of zero. The average field
         strength in both cases is $\Bav=2.4\times 10^{17}$ G.}
\end{figure}

Purely toroidal fields give prolate density distributions, although we
find that the surface shape remains virtually spherical even for large
ellipticities (i.e. strong fields). Because rotation gives rise to
oblateness in stars, it opposes the effect of a toroidal field in a
star, and the two effects can balance to give a rotating magnetised
star with zero overall ellipticity. Note that in this case the stars
will have oblate surface shapes but a spherical density distribution
--- see figure \ref{tordens}.

\begin{figure}
\begin{minipage}[c]{\linewidth}
\begin{center}
\psfrag{B}{$\displaystyle{\frac{\Bav}{10^{17}\textrm{ G}}}$}
\psfrag{Om_K}{$\displaystyle{\frac{\Omega_K}{2\pi}}$ / Hz}
\includegraphics[width=0.4\linewidth]{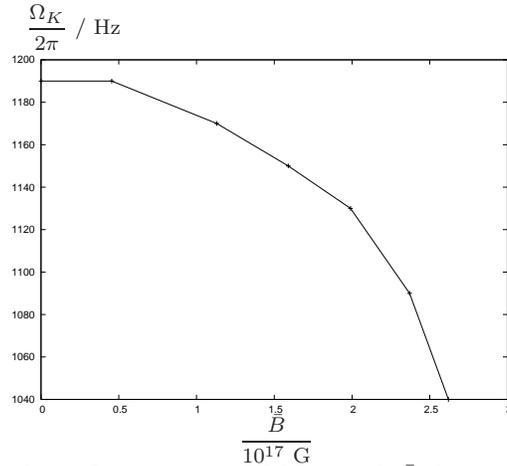}
\end{center}
\end{minipage}
\caption{\label{BvsOmK}
         The dependence of Keplerian velocity $\Omega_K$ on magnetic field
         strength $\Bav$, for stars with purely poloidal fields. Note
         that an appreciable decrease in $\Omega_K$ only occurs for very
         strong fields.}
\end{figure}

Next we look at the effect of magnetic fields on the Keplerian
velocity $\Omega_K$ --- see figure \ref{BvsOmK}. We find that whilst
increasing the field strength
causes a slight decrease in the velocity needed to cause mass
shedding, this effect only becomes noticeable for very strong
fields. It seems, therefore, that magnetic fields are unlikely to
affect the stability of a star in this manner.

\begin{figure}
\begin{minipage}[c]{\linewidth}
\begin{center}
\psfrag{(a)}{(a)}
\psfrag{(b)}{(b)}
\psfrag{(c)}{(c)}
\psfrag{(d)}{(d)}
\includegraphics[width=0.7\linewidth]{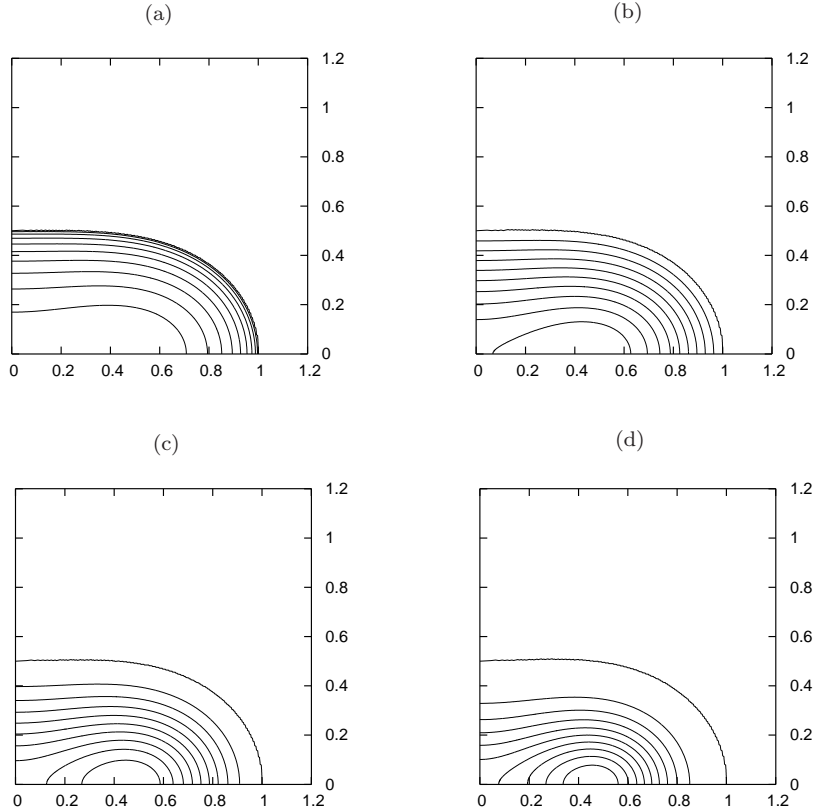}
\end{center}
\end{minipage}
\caption{\label{varypoly}
         Non-rotating configurations, all with a purely poloidal field
         and an axis ratio of $r_p/r_{eq}=0.5$. Plots (a) to (d)
         are for $N=0.5,1,1.5,2$ polytropes, respectively; the
         corresponding field strengths are $\Bav=7.62,4.31,2.98,1.13\
         \times 10^{17}$ G, the maximum densities are
         $1.67,1.14,0.623,0.179 \times 10^{15}$ g cm${}^{-3}$ and the
         equatorial radii are $r_{eq}=10.2,12.9,17.6,29.6$ km, respectively.}  
\end{figure}

We have generally presented results for an $N=1$ polytrope, as this is
regarded as a reasonable approximation to a neutron star. For our
final two figures, however, we briefly investigate the effect of varying
the polytropic index $N$, whilst maintaining a mass of $1.4M_\odot$ and
central density of $2.19\times 10^{15}$ g cm${}^{-3}$ in the
corresponding unmagnetised `background' polytropic star. In
figure \ref{varypoly}, we plot four stars with the same surface
distortion $r_p/r_{eq}=0.5$ but different $N$. We see that when $N$ is 
low the density contours are all close to the edge of the star, with a
large (slightly off-centre) high-density region; in the
limiting case $N=0$ the star is an incompressible, uniform density
configuration, so all contour lines coincide with the star's
surface. For higher values of $N$ the high-density region
becomes smaller and the low-density outer region becomes larger. We
note that the $N=2$ polytrope shown cannot be a neutron star model,
however, as its maximum density of $1.79 \times 10^{14}$ g cm${}^{-3}$
is lower than the density of heavy nuclei,
$\rho_0 = 2.4 \times 10^{14}$ g cm${}^{-3}$.

\begin{figure}
\begin{minipage}[c]{\linewidth}
\begin{center}
\psfrag{N}{$N$}
\psfrag{Bav}{$\displaystyle{\frac{\Bav}{10^{17}\textrm{ G}}}$}
\includegraphics[width=0.4\linewidth]{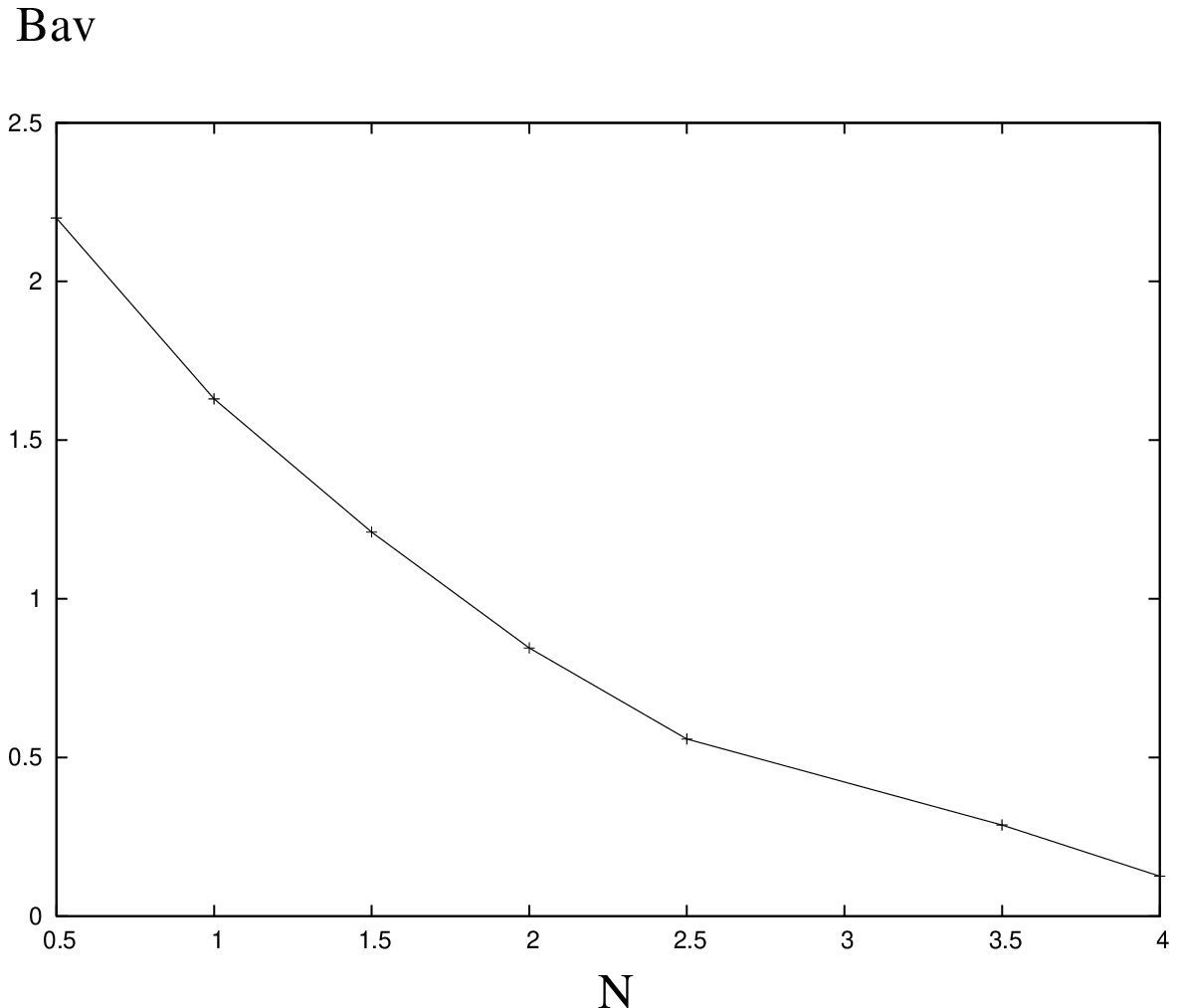}
\end{center}
\end{minipage}
\caption{\label{NvsB}
         Right: the poloidal field strength $\Bav$ required to induce
         a surface distortion of $r_p/r_{eq}=0.95$, plotted for
         various polytropic indices. We see that the required field is
         weaker for higher-$N$ polytropes.}
\end{figure}

Finally, in figure \ref{NvsB}, we look at non-rotating stars
magnetised by a purely poloidal field, with an axis ratio of $0.95$. We
plot the dependence of the field strength on polytropic index $N$,
finding that as $N$ is increased a weaker field is required to support
the same surface distortion.

\section{Discussion}

To understand how strong magnetic distortions may be in highly
magnetised objects like magnetars, realistic models are needed to
study the field structure of these stars. The
formalism we use in this work comes directly from the assumptions of
axisymmetry and perfect conductivity, together with a boundary
condition that the poloidal part of the field should decay at infinity
rather than vanishing at the star's surface; we anticipate that these
conditions provide a reasonable model of a neutron star's field.

The general formalism of axisymmetric MHD reduces to a mixed-field
case and a purely toroidal-field case, with two (mathematically)
arbitrary functions in the former case ($\kappa(u)$ and $\alpha(u)$)
and one in the latter ($h(\gamma)$). Despite the apparent freedom in
choosing these functions, we found that on physical grounds only one
functional form was satisfactory for each one; see section
\ref{mag-restrict}. We conclude that the equations we have numerically
solved in this work are in fact quite general and that we have not
excluded physically valid branches of solutions with our choices.

Perturbative calculations in the weak-field regime have found that
$\eps$ depends linearly on $\Bav^2$. With the use of our nonlinear code we
are able to investigate how well this approximation holds for larger
fields and ellipticities. We can see graphically that the first few
points from both plots in figure \ref{ellipvsB2} lie in fairly
straight lines and hence we deduce the relations
\beq \label{epspol}
\eps_{pol}\sim 5\times 10^{-4} \brac{\frac{\Bav}{10^{16}\textrm{ G}}}^2
          \sim 2\times 10^{-3} \brac{\frac{B_p}{10^{16}\textrm{ G}}}^2
\eeq
for the purely poloidal case (the above relation also uses
$B_p/\Bav\sim 0.5$ from figure \ref{N-Bratio}), and
\beq \label{epstor}
\eps_{tor}\sim -3\times 10^{-4} \brac{\frac{\Bav}{10^{16}\textrm{ G}}}^2
\eeq
for the purely toroidal case;
where in both cases we have used a star of mass $1.4M_\odot$ whose
radius would be 10 km if unmagnetised.
By comparing these extrapolated linear-regime formulae with our
non-linear code results, we can explore how well perturbative results
are likely to hold in a strong-field regime. We find that the linear
regime given by \eqref{epspol} and \eqref{epstor} differs by less than
10\% of the actual non-linear code result (shown in figure
\ref{ellipvsB2}) provided that $\Bav\lesssim 1.5\times 10^{17}$
G, or equivalently $\eps\lesssim 0.15$. Alternatively, if we allow the
linear relation to differ by up to 30\% from the nonlinear result, we
may use the linear relation as an `acceptable' approximation for
$\Bav\lesssim 3\times 10^{17}$ G or $\eps\lesssim 0.35$ (i.e. it holds
for the entire range of ellipticities we can plot in the
toroidal-field case).

This suggests that for
\emph{all} known neutron star field strengths, $\eps$ is likely to be
linearly dependent on $\Bav^2$, to a good approximation. Hence
perturbation theory \emph{could} provide accurate predictions of NS
distortions, provided the neutron star model used is also a close
approximation to real NS physics.

We are also able to compare our linear-regime formulae with
the analytic work of \citet{haskell}, who also treated pure poloidal
fields extending outside the star and pure toroidal fields vanishing
at the stellar surface (as for our work). For the same mass, radius and
polytropic index their formulae give:
\beq
\eps_{pol}\sim 10^{-2} \brac{\frac{B_s}{10^{16}\textrm{ G}}}^2
\ \textrm{ and }\  
\eps_{tor}\sim -2\times 10^{-4} \brac{\frac{\Bav}{10^{16}\textrm{ G}}}^2
\eeq
where $B_s$ is the surface magnetic field strength, which was assumed
constant for the calculation of \citet{haskell}; we do not have a
constant surface field so have compared with their work using the
value of $|{\bf{B}}|$ at the pole instead. Since their field
geometries are clearly not identical to ours, and since we had to
extrapolate to obtain our formulae, we would not expect precise
agreement. Nonetheless, we feel that the similarities show that our
work makes sensible contact with perturbative calculations.

From figure \ref{ellipvsB2}, beginning at an unmagnetised spherical
star, we find that in both the poloidal and toroidal cases the
magnetic field strength required increases for larger
distortions, initially; as would be expected from perturbative
work. However, in the purely poloidal case the field strength then
\emph{peaks} at $\eps\sim 0.8$, dropping slightly as $\eps$ is increased
further. Around the same point the density distribution becomes
toroidal in nature --- that is, the point of maximum density moves
away from the centre and a high-density torus forms; this leads us to
speculate that at $\eps\sim 0.8$ it
becomes energetically favourable for the density to change from a
spheroidal profile (as seen in the weaker-field stars, e.g. figure
\ref{torus-shape}, plot (a)) to a toroidal one (e.g. figure
\ref{torus-shape}, plot (d)). It is clear that if the magnetic
field in a star is increased beyond the peak value of
$\sim 5\times 10^{17}$ G shown in the left-hand plot of figure
\ref{ellipvsB2} then one of our initial assumptions must be violated.
Since we cannot investigate the possibilities with our current code,
we conclude that a hypothetical star with a field of $\Bav > 6\times
10^{17}$ may either have no equilibrium solution (in which case it may
lose magnetic energy until it is in equilibrium), or that there may be
a new triaxial branch of super-magnetised solutions bifurcating from
the biaxial curve at $\eps\sim 0.8$.

We do not find a similar peaking of the field strength in the purely
toroidal case, however. In this case the largest ellipticities we are
able to calculate are around $\eps\sim 0.35$. Whilst this particular
value may represent a limitation of our numerical scheme, we suggest
that a limited range of ellipticities is a consequence of the formalism
for toroidal fields in axisymmetry, where ${\bf B}$ is directly linked
to the density $\rho$; in the mixed-field case we have a separate
equation to iteratively solve for the magnetic field. Thus
restrictions on the field geometry may restrict the size of
permissible ellipticities.

Of course, whilst the `peak field strength' we discuss here is a
theoretical upper bound on NS fields, there are probably other
physical effects that place a
lower bound than $\sim 5\times 10^{17}$ G on the maximum
field. Certainly, if magnetar surface fields are $\sim 10^{15}$ G one
would not expect their volume-averaged fields to exceed $\sim 10^{16}$
G.

We have argued that the equations we solve in this paper lead to quite
general solutions for axisymmetric stars. However, we find that
although it is possible to find solutions with purely poloidal or
purely toroidal fields, the range of mixed-field solutions is very
limited. Using $\etor/\emag$ as a gauge of the strength of the
toroidal component in a mixed-field star, we find that for all our
stars $0 \leq \etor/\emag < 0.07$. The other extreme is of course
$\etor/\emag=1$ for purely toroidal fields. This means that although
the toroidal component does have some influence in a mixed-field star
(see table at the end of section \ref{magfields}), it is dominated by
the effect of the poloidal field. In particular all our mixed-field
stars have oblate density distributions.

Our mixed-field stars have the boundary condition that the toroidal
component vanishes at the surface, whilst the poloidal piece only
decays at infinity. By contrast, \citet{haskell} considered the
problem of mixed-field stars where the total field vanished at the
surface. This results in an eigenvalue problem, with all (discrete)
solutions having prolate density distributions. Since the chief
difference between our work appears to be the choice of boundary
condition, we speculate that our boundary condition favours poloidal
distortions, whilst that of \citet{haskell} favours the toroidal
component.

The numerical simulations of \citet{braithtorpol} suggest that a
stable magnetic field will have $0.20\lesssim\etor/\emag\lesssim
0.95$. If this result is directly applicable to our work then it would
imply that \emph{none} of the solutions that exist within our
axisymmetric formalism are stable. However, for numerical reasons
these simulations use a magnetic diffusivity term which is zero within
the star and increases through a transition region to a high, constant
value in the exterior (see \cite{braithwaite} for details). We suggest
that this transition region may favour the toroidal component of a
mixed-field star; it would be interesting to see if a similar
stability result emerges from simluations using a boundary condition
more similar to ours.

Although we regard our boundary condition as the most natural for a
mixed-field fluid with infinite conductivity, neutron stars are not
perfect conductors. In moving from the superfluid
interior to the crust and magnetosphere, it is clear that the
resistivity of the medium increases and hence the boundary condition
should be adapted to reflect this. \citet{colaiuda} noted this and
attempted to mimic more `natural' boundary behaviour by allowing the
poloidal part of the field to extend outside the star (as for our
field), but matching the toroidal part to a surface current rather than
forcing it to vanish at the surface.

With no clear idea about the nature of currents on the surface of
neutron stars, we suggest that it may be easier to neglect their
effects, so that the toroidal-field component vanishes at the surface.
Incorporating the effects of resistivity in the outer regions of the
neutron star would then involve adapting the boundary condition for the
poloidal component; this could resemble a surface treatment somewhere
between ours (where the poloidal field is
unaffected by passing through the surface) and that of \citet{haskell}
(where the poloidal field decays at the surface). Since our boundary
condition gives a poloidal-dominated field and that of \citet{haskell}
gives a toroidal-dominated field, we suggest that the inclusion of
resistivity would result in configurations where neither component is
universally dominant. In particular, we would not expect magnetic
distortions in real, mixed-field, neutron stars to be universally
oblate or prolate. We conclude that future, more realistic, models of
magnetised stars should incorporate a boundary condition like ours,
but modified to take account of the increasing resistivity in the
outer regions of the neutron star.

\section*{Acknowledgments}

The magnetic code used to generate our results was based on a code for
purely rotating stars written by Nikolaos Stergioulas. We thank Nils
Andersson and Brynmor Haskell for helpful discussions. We are also
grateful to the anonymous referee for their careful reading of this
paper and useful comments. This work was supported by STFC through
grant number PP/E001025/1.

\appendix

\section{Axisymmetric magnetic fields}

\subsection{General forms for magnetic field and current}
\label{genmagderivation}

We wish to see how the assumption of axisymmetry constrains the
geometry of the magnetic field and the current; and hence also the
form of the Lorentz force.  Working in cylindrical
polar coordinates, we begin with the equilibrium equation for a
magnetised rotating fluid:
\beq \label{divnewton2}
-\nabla H - \nabla\Phi + \nabla\brac{\frac{\Omega^2\pom^2}{2}}
          + \frac{\lor}{\rho} = 0
\eeq 
where we have rewritten \eqref{newton2} above by replacing the usual
$\nabla P/\rho$ term with the gradient of the enthalpy $H=\int_0^P
dP'/\rho(P')$ and also explicitly written the centrifugal
term as the gradient of a scalar.

If we now take the curl of \eqref{divnewton2} then by the vector
identity $\curl\nabla f =0$ (for any scalar field $f$) we see that
\beq
\curl\brac{\frac{\lor}{\rho}} = 0,
\eeq
implying that $\lor/\rho$ is also the gradient of some scalar
$M$. Note that $\nabla M \cdot \bB=0$, i.e. $M$ is constant along
field lines.

Next we write $\bB$ in terms of a streamfunction $u$, defined through
the relations
\beq
B_\pom = -\frac{1}{\pom}\pd{u}{z}\ ,\ B_z = \frac{1}{\pom}\pd{u}{\pom}
\eeq
--- note that these components give a solenoidal magnetic field,
$\div\bB=0$, by construction.
Hence
\beq
\bB = -\frac{1}{\pom}\pd{u}{z}\be_\pom + B_\phi \be_\phi
      +\frac{1}{\pom}\pd{u}{\pom}\be_z.
\eeq
Now comparing the equation with
\beq
\nabla u\times\be_\phi = -\pd{u}{z}\be_\pom+\pd{u}{\pom}\be_z,
\eeq
we see that $\bB$ may be written as
\beq \label{axifield}
\bB = \frac{1}{\pom}\nabla u\times\be_\phi + B_\phi \be_\phi.
\eeq
Note that this implies $\bB\cdot\nabla u=0$, i.e. $u$ is constant
along field lines. Recalling that $M$ also has this property, we
deduce that
\beq
M = M(u).
\eeq

Next we turn to Amp\`ere's law in axisymmetry:
\beq
4\pi\bj = \curl\bB = -\pd{B_\phi}{z}\be_\pom
                     +\brac{\pd{B_\pom}{z}-\pd{B_z}{\pom}}\be_\phi
                     +\frac{1}{\pom}\pd{ }{\pom}(\pom B_\phi)\be_z.
\eeq
Now by comparing the poloidal part of the current
\beq
\bj_{pol}= -\frac{1}{4\pi\pom}\pd{ }{z}(\pom B_\phi)\be_\pom
           +\frac{1}{4\pi\pom}\pd{ }{\pom}(\pom B_\phi)\be_z
\eeq
with the quantity
\beq
\nabla(\pom B_\phi)\times\be_\phi =
           -\pd{ }{z}(\pom B_\phi)\be_\pom
           +\pd{ }{\pom}(\pom B_\phi)\be_z,
\eeq
we see that
\beq \label{axjpol}
\bj_{pol}=\frac{1}{4\pi\pom}\nabla(\pom B_\phi)\times\be_\phi.
\eeq
Next we consider the toroidal part of the current
$\bj_{tor}=j_\phi\be_\phi$ and rewrite $j_\phi$ using the definition
of the streamfunction $u$:
\beq \label{axjtor}
4\pi j_\phi = \pd{B_\pom}{z}-\pd{B_z}{\pom}
            = -\frac{1}{\pom}\brac{
                   \pom\pd{ }{\pom}\brac{\frac{1}{\pom}\pd{u}{\pom}}
                   +\pd{^2 u}{z^2} }.
\eeq
For brevity we define a differential operator $\Delta_*$ by
\beq
\Delta_* \equiv \pd{^2}{\pom^2} - \frac{1}{\pom}\pd{ }{\pom}
                                  + \pd{^2}{z^2}.
\eeq
Now using this definition together with \eqref{axjpol} and
\eqref{axjtor} we see that the current may be written as
\beq \label{axicurrent}
4\pi\bj = \frac{1}{\pom}\nabla(\pom B_\phi)\times\be_\phi
          -\frac{1}{\pom}\GSu\ \be_\phi.
\eeq

Our two key results from this section so far are the expressions
\eqref{axifield} and \eqref{axicurrent} for the general form of an
axisymmetric magnetic field and current, respectively. Next we
consider the form of the Lorentz force arising from these two
quantities. We see that in general
\beqy \nn
\lor = \bj\times\bB 
    &=& \brac{\bj_{pol}+j_\phi\be_\phi}
        \times \brac{\bB_{pol}+B_\phi\be_\phi}\\
    &=& \underbrace{   \bj_{pol}\times\bB_{pol}   }_{\lor_{tor}}
        +\underbrace{  j_\phi\be_\phi\times\bB_{pol}
                      +B_\phi\bj_{pol}\times\be_\phi  }_{\lor_{pol}}.
\label{lortorpol}
\eeqy
Returning to our original force balance equation \eqref{divnewton2} we
note that the pressure, gravitational and centrifugal forces are
axisymmetric (i.e. no $\phi$-dependence); therefore $\lor$ is also
axisymmetric and its toroidal component must vanish:
\beq \label{lortorzero}
\lor_{tor} = \bj_{pol}\times\bB_{pol} = 0.
\eeq
At this point there are two ways to proceed: either $\bB_{pol}$ is
non-zero, in which case $\bB_{pol}$ and $\bj_{pol}$ are parallel; or
$\bB_{pol}=0$. We shall consider these cases separately in the next two
subsections.

\subsection{Mixed poloidal and toroidal fields; the Grad-Shafranov equation}
\label{mix_formal}

We have shown that the requirement \eqref{lortorzero} follows from the
axisymmetry of our problem. In this subsection we consider the case
where $\bB_{pol}$ and $\bj_{pol}$ are parallel, corresponding to a
magnetic field with both poloidal and toroidal components. We will see
that the form of purely poloidal magnetic fields may be found as a
special case of the general mixed-field configuration.

Recall from \eqref{axifield} and \eqref{axjpol} that
\beqy \nn
\bB_{pol} &=& \frac{1}{\pom}\nabla u\times\be_\phi \\
\bj_{pol} &=& \frac{1}{4\pi\pom}\nabla(\pom B_\phi)\times\be_\phi.
\nn
\eeqy
Knowing that these two quantities are parallel we see that $u$ and
$\pom B_\phi$ must be related by some function $f$:
\beq
\pom B_\phi = f(u).
\eeq

Next we evaluate the non-zero Lorentz force components,
i.e. $\lor_{pol}$ from \eqref{lortorpol}. Using the pair of equations
at the start of this subsection, we find that
\beq
\be_\phi\times\bB_{pol}
       =\be_\phi\times\brac{\frac{1}{\pom}\nabla u\times \be_\phi}
       =\frac{1}{\pom}\brac{\nabla u - \be_\phi(\be_\phi\cdot\nabla u)}
       =\frac{1}{\pom}\nabla u
\eeq
and similarly
\beq
\bj_{pol}\times\be_\phi = -\frac{1}{4\pi\pom} \nabla (\pom B_\phi).
\eeq
Now using these expressions in \eqref{lortorpol}, together with the
relation $4\pi j_\phi=-\frac{1}{\pom}\GSu$ from \eqref{axicurrent}, we find
that
\beq
4\pi\lor = -\frac{1}{\pom^2}\GSu\ \nabla u
           -\frac{1}{\pom} B_\phi \nabla(\pom B_\phi)
\eeq
which, recalling the definitions $\nabla M = \lor/\rho$ and
$f(u)=\pom B_\phi$, becomes
\beq
4\pi\rho\nabla M = -\frac{1}{\pom^2}\GSu\ \nabla u
                   -\frac{1}{\pom^2} f(u) \nabla f(u).
\eeq
Since $M$ and $f$ are both functions of $u$ alone we are able to
rewrite $\nabla M(u)$ and $\nabla f(u)$ using the chain rule, to give
\beq
-4\pi\rho\td{M}{u}\nabla u = \frac{1}{\pom^2}\GSu\ \nabla u
                            +\frac{1}{\pom^2}f(u)\td{f}{u}\nabla u.
\eeq
Now provided $\nabla u \neq 0$ we have
\beq \label{gradshafranov}
4\pi\rho\td{M}{u} = -\frac{1}{\pom^2}
                            \brac{\GSu + f(u)\td{f}{u}},
\eeq
which is the Grad-Shafranov equation \citep{gradrubin}.

We now return to the general form of an axisymmetric current
\eqref{axicurrent}, replacing $\pom B_\phi$ with $f(u)$ and using the
chain rule to give:
\beq \label{torpolj1}
4\pi\bj = \frac{1}{\pom}\td{f}{u}\nabla u\times\be_\phi
         -\frac{1}{\pom}\GSu \be_\phi.
\eeq
We now use \eqref{axifield} to make the replacement
$\frac{1}{\pom}\nabla u\times\be_\phi=\bB_{pol}$ and the
Grad-Shafranov equation \eqref{gradshafranov} to eliminate $\GSu$ from
\eqref{torpolj1}:
\beq
4\pi\bj = \td{f}{u}\bB_{pol}
         +\frac{1}{\pom}\brac{4\pi\pom^2\rho\td{M}{u}+f(u)\td{f}{u}}.
\eeq
Finally we use the definition $f=\pom B_\phi$ and
$\bB = \bB_{pol}+B_\phi \be_\phi$ to yield an expression for the
current in terms of the magnetic field and the derivatives of the
functions $M(u)$ and $f(u)$:
\beq \label{mixedjwithfM}
\bj = \frac{1}{4\pi}\td{f}{u}\bB + \rho\pom\td{M}{u}\be_\phi.
\eeq

\subsection{Purely poloidal field}

Having arrived at an expression for an axisymmetric current associated
with a mixed poloidal-toroidal field \eqref{mixedjwithfM}, we may
straightforwardly specialise to purely poloidal magnetic fields by
choosing $f(u)$ as a constant. Then $\td{f}{u}=0$ and the mixed term
vanishes from the expression for $\bj$, leaving only a toroidal
current
\beq \label{poljwithM}
\bj = \rho\pom\td{M}{u}\be_\phi
\eeq
and hence a purely \emph{poloidal} field, by Amp\`ere's law.

\subsection{Purely toroidal field}
\label{tor_formal}

In the previous subsection we showed that \eqref{mixedjwithfM} may be
trivially reduced to the poloidal-field case. However it is clear from
the form of \eqref{mixedjwithfM} that there is no choice of $f$ and
$M$ which yields a poloidal current (or equivalently a toroidal
field). Setting $M(u)$ to be a constant, for example, results in the
general expression for a force-free field
\beq
\bj = \frac{1}{4\pi}\td{f}{u}\bB,
\eeq
which is of less interest to us, as we aim to study distortions caused
by magnetic fields.

It is clear that the derivation used for mixed fields does not hold in
the toroidal-field case. Previously we were able to use
\eqref{lortorzero} to simplify the current-field relation, but no such
constraint is provided for a toroidal field, where
$\bB_{pol}=0$. Accordingly we must return to subsection
\ref{genmagderivation} where we found that
\beqy
\bB_{pol} &=& \frac{1}{\pom}\nabla u\times\be_\phi \nn\\
\bj_{pol} &=& \frac{1}{4\pi\pom}\nabla(\pom B_\phi)\times\be_\phi \nn
\eeqy
(from equations \eqref{axifield} and \eqref{axjpol}).
Since $\bB_{pol}=0$ we no longer require $\pom
B_\phi$ to be a function of $u$; indeed the streamfunction $u$ will
not even enter our final solution. We also recall that the general
form of an axisymmetric Lorentz force is given by \eqref{lortorpol},
which in the case of $\bB_{pol}=0$ reduces to
\beq
\lor = B_\phi\bj_{pol}\times\be_\phi.
\eeq
Using \eqref{axjpol} to replace $\bj_{pol}$ in this expression then
gives
\beq \label{lortor1}
\lor=\frac{B_\phi}{4\pi\pom}\brac{\nabla(\pom B_\phi)\times\be_\phi}
       \times\be_\phi
    =-\frac{B_\phi}{4\pi\pom}\nabla(\pom B_\phi).
\eeq
Again recalling previous work in this section, we note that taking the
curl of \eqref{divnewton2} shows that $\curl(\lor/\rho)=0$. We use
this fact together with the vector identity
$\curl(f\nabla g)=\nabla f\times\nabla g$ to rewrite \eqref{lortor1}
as
\beq
\nabla\brac{\frac{B_\phi}{\rho\pom}}\times\nabla(\pom B_\phi) = 0.
\eeq
If we write $\frac{B_\phi}{\rho\pom}$ in the above expression as
$\frac{1}{\rho\pom^2}\pom B_\phi$ and use the chain rule, some algebra
leads to
\beq
-\frac{B_\phi}{\rho^2\pom^3}\nabla(\rho\pom^2)
                           \times\nabla(\pom B_\phi) = 0.
\eeq
Provided $B_\phi/\rho^2\pom^3\neq 0$ we then deduce that
$\nabla(\rho\pom^2)\times\nabla(\pom B_\phi)=0$ and hence that
$\rho\pom^2$ and $\pom B_\phi$ are related by some function $h$, i.e.
\beq \label{Bphitor}
\pom B_\phi = h(\rho\pom^2).
\eeq
As before we now define a magnetic function $M$ through $\lor/\rho=\nabla
M$ (note that here $M$ need not be a function of the streamfunction $u$ of
previous sections). From \eqref{lortor1} and \eqref{Bphitor} we then
find that
\beq
\nabla M = -\frac{h(\rho\pom^2)}{4\pi\rho\pom^2}\nabla h(\rho\pom^2).
\eeq
By the chain rule we have
$\nabla h(\gamma)=\td{h}{\gamma}\nabla\gamma$ where we have introduced
the notation $\gamma\equiv\rho\pom^2$. Given this we have
\beq
\nabla M = -\frac{h(\gamma)}{4\pi\gamma}\td{h}{\gamma}\nabla\gamma
\eeq
and so
\beq
M = -\frac{1}{4\pi}\int_0^{\rho\pom^2}
                      \frac{h}{\gamma}\td{h}{\gamma}\ d\gamma.
\eeq

\label{lastpage}

\end{document}